\documentclass[twoside]{dis04}

\def\runauthor{Wu-Ki Tung}
\def\shorttitle{Status of Global QCD analysis of PDF's.}

\newenvironment{Mylis}[1][]
{\begin{list}
 {#1}
 {\settowidth{\labelwidth}{#1}\setlength{\labelsep}{0.5em}
  \setlength{\leftmargin}{\labelwidth}\addtolength{\leftmargin}{\labelsep}
  \setlength{\rightmargin}{0em}
  \setlength{\parsep}{0.3\parskip}\setlength{\itemsep}{0.3\parskip}
  \setlength{\topsep}{0ex}}
}
{\end{list}}

\newcommand{\mylis}[2][]
{\begin{Mylis}[#1] #2 \end{Mylis}}

\newcommand{\Table}{
\begin{table}[ht]
\centering  \resizebox*{0.8\textwidth}{!}{\includegraphics{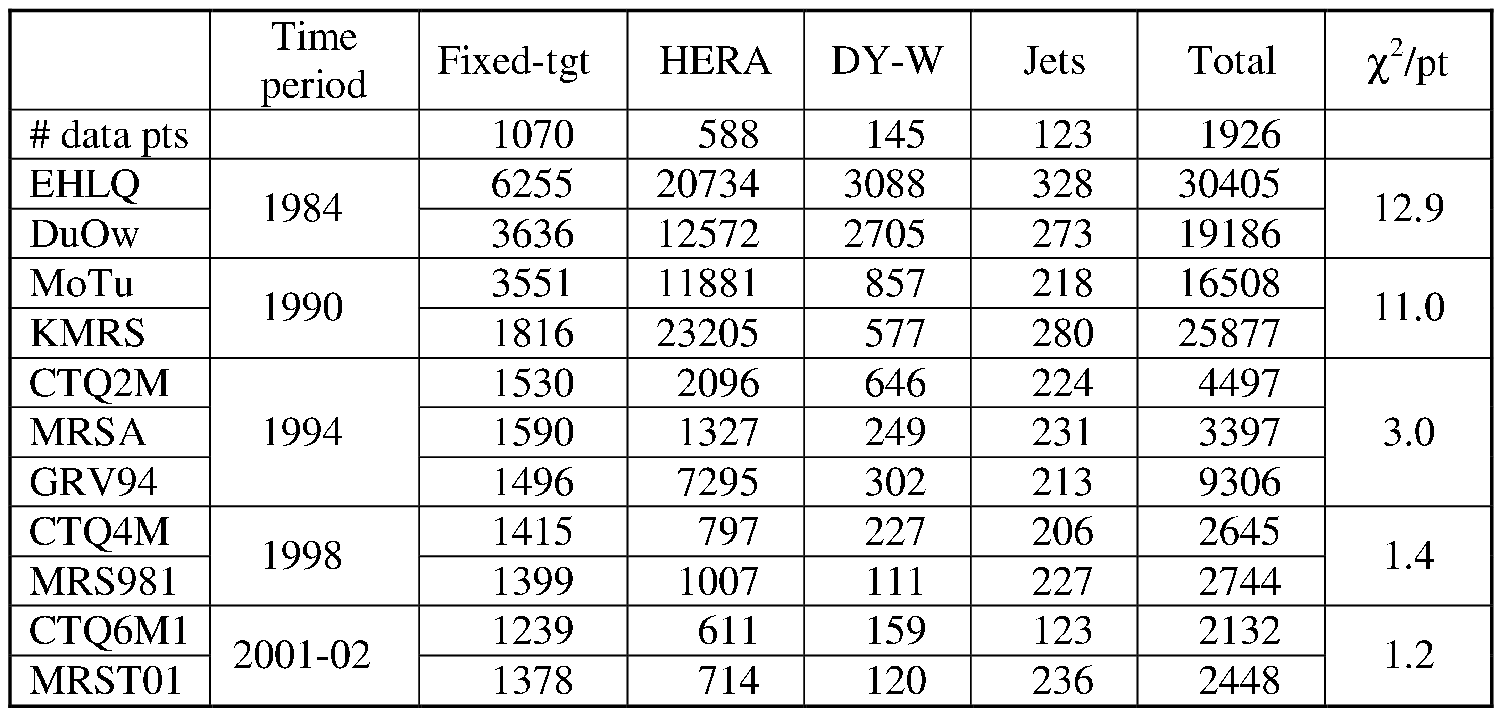} }
\caption{Historical PDFs evaluated on current data sets in the
{\protect\scriptsize {$\overline{MS}$}\ scheme, to show the steady progress
that has been made in determining the parton structure of the nucleon. Focus
is on the historic trend, particularly illustrated by the last column,
rather than on detailed numbers in neighboring rows, which are subject to
various caveats. (See text.) }}
\label{tbl:chi}
\end{table}
}
\newcommand{\udquarks}
{
\begin{figure}[ht]
\resizebox{0.49\textwidth}{!}{\includegraphics{figs/Uqk1.eps}} \hfill %
\resizebox{0.49\textwidth}{!}{\includegraphics{figs/Dqk1.eps}}
 \caption{Historical evolution of the u quark (a) and d quark (b) distributions}
\label{fig:udquarks}
\vspace{-3ex}
\end{figure}
}
\newcommand{\GluAB}{
\begin{figure}[ht]
\resizebox{0.49\textwidth}{!}{\includegraphics{figs/GluA.eps}} \hfill %
\resizebox{0.49\textwidth}{!}{\includegraphics{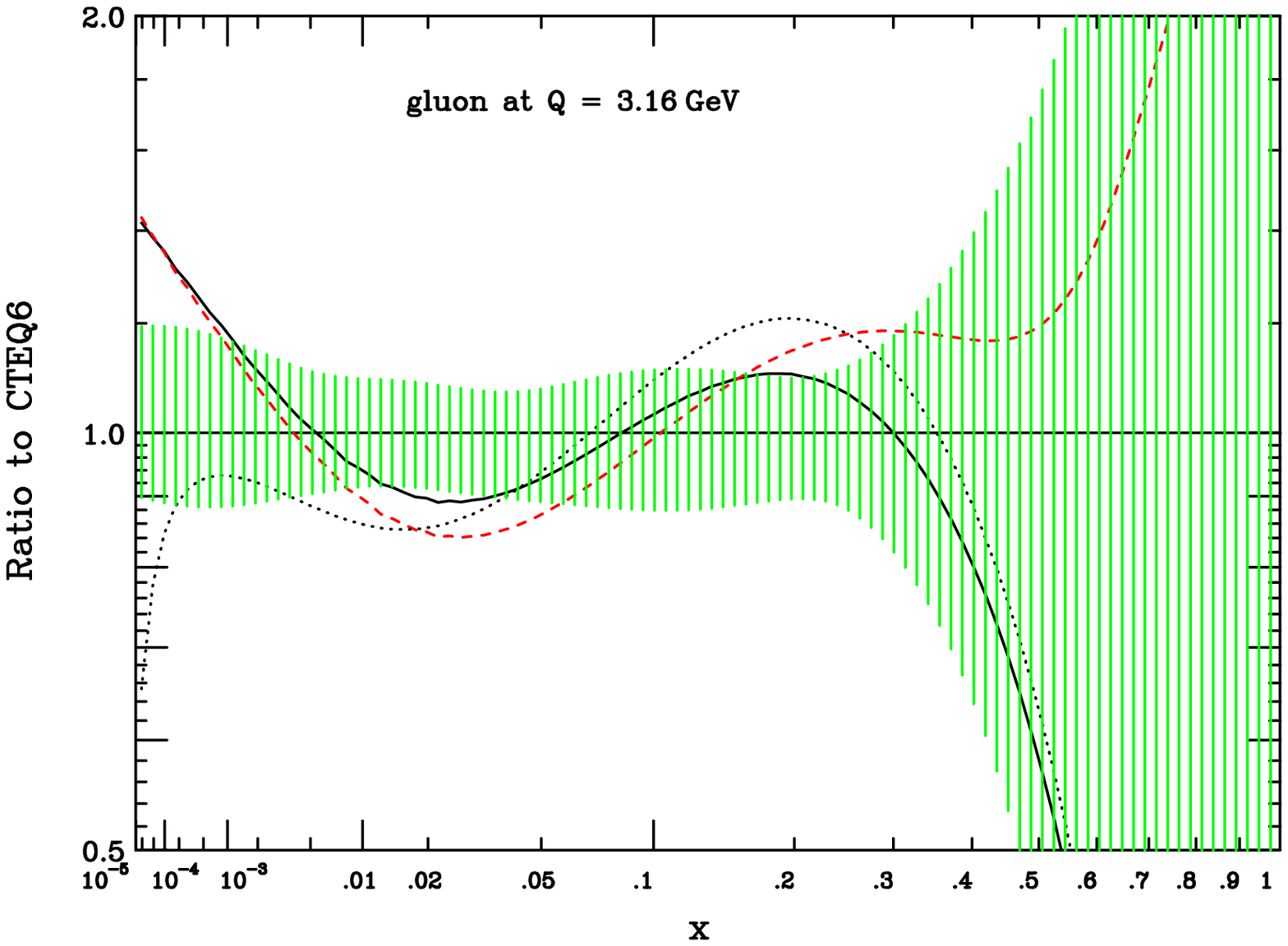}}
\caption{(a) Historical evolution of the gluon distribution; and (b) current
range of uncertainty of the gluon distribution estimated by CTEQ.}
\label{fig:GluAB}
\end{figure}
}
\newcommand{\UdAsym}{
\begin{figure}[ht]
\resizebox{0.49\textwidth}{!}{\includegraphics{figs/DoU.eps}} \hfill
\resizebox{0.49\textwidth}{!}{\includegraphics{figs/UdAsym.eps}}
\caption{(a) The $d(x,Q)$ to $u(x,Q)$ ratio as a function of $x$; and
(b) The asymmetry between the non-strange light sea quarks $\bar{d}$
and $\bar{u}$.}
\label{fig:UdAsym}
\end{figure}
}
\newcommand{\StrAsym}{
\begin{figure}[ht]
\resizebox{0.49\textwidth}{!}{\includegraphics{figs/SoUd.eps}} \hfill
\resizebox{0.49\textwidth}{!}{\includegraphics{figs/CompNumAsym.eps}}
\caption{(a) The asymmetry between the strange and the light non-strange
sea quarks; and (b) The strangeness asymmetry as a function of $x$: shown
are results from a recent global analysis, along with two earlier results
cf.\ \protect\cite{StrAsym} for explanation.}
\label{fig:StrAsym}
\vspace{-3ex}
\end{figure}
}
\newcommand{\charm}{
\begin{figure}[ht]
\centerline{
\resizebox{0.49\textwidth}{!}{\includegraphics{figs/chm1.eps}}
}
\caption{(a) The charm distribution function.}
\label{fig:charm}
\vspace{-3ex}
\end{figure}
}
\newcommand{\wXsecErr}{
\begin{figure}[ht]
\vspace{-3 ex}
\centerline{ \resizebox{0.8\textwidth}{!}{\includegraphics{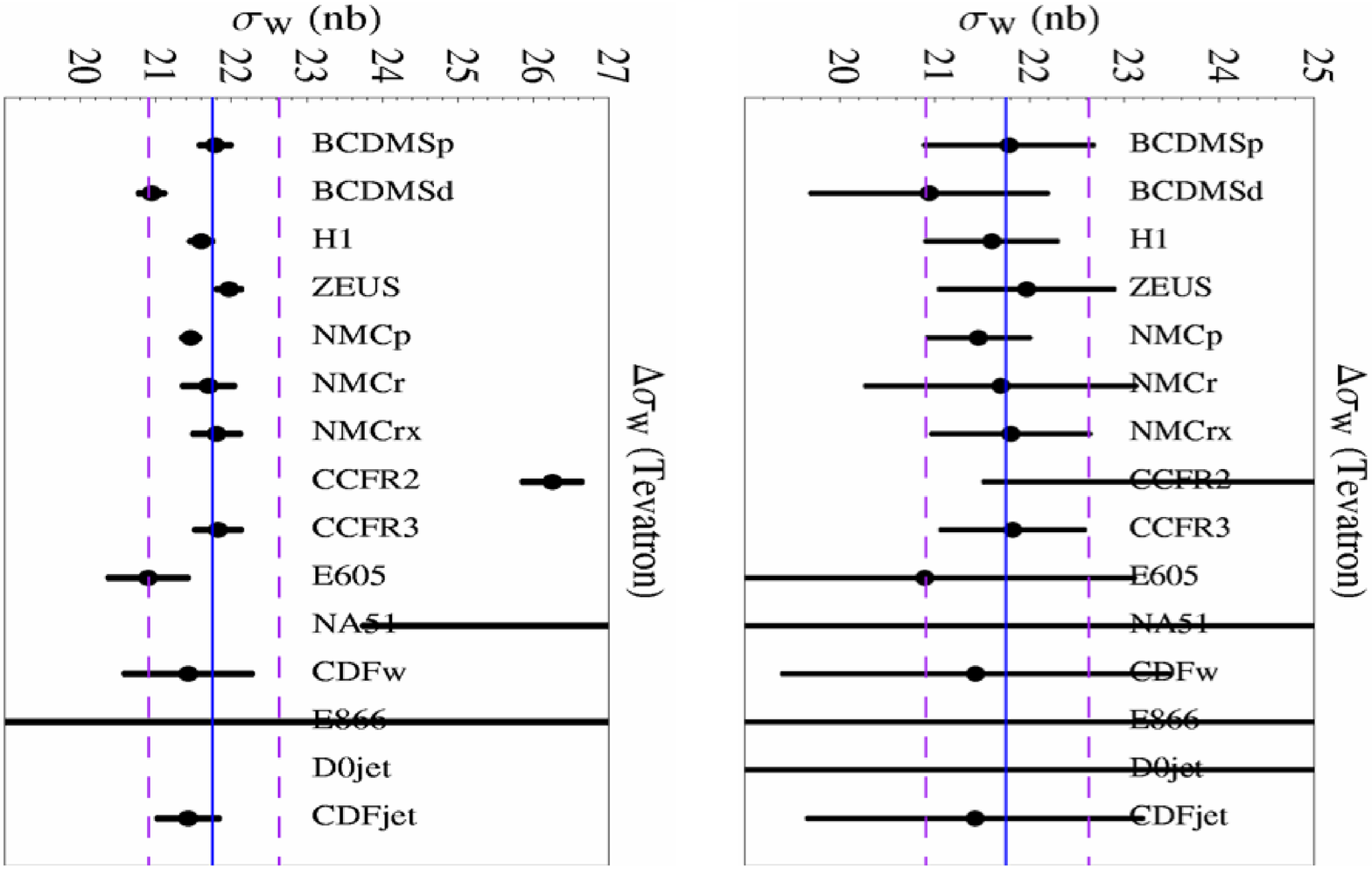}}
}
\caption{Predicted value of $\sigma_W$ at the Tevatron: (a) $\Delta\chi^2=1$ error ranges for
individual experimental data sets evaluated from
PDF sets obtained by the Lagrange Multiplier method in constrained global fits; and
(b) 90\% confidence level ranges for the same data sets and PDF sets.}
\label{fig:wXsecErr}
\vspace{-4ex}
\end{figure}
}
\newcommand{\Unc}{
\begin{figure}[ht]
\vspace{-3ex}
\resizebox{0.50\textwidth}{!}{\includegraphics{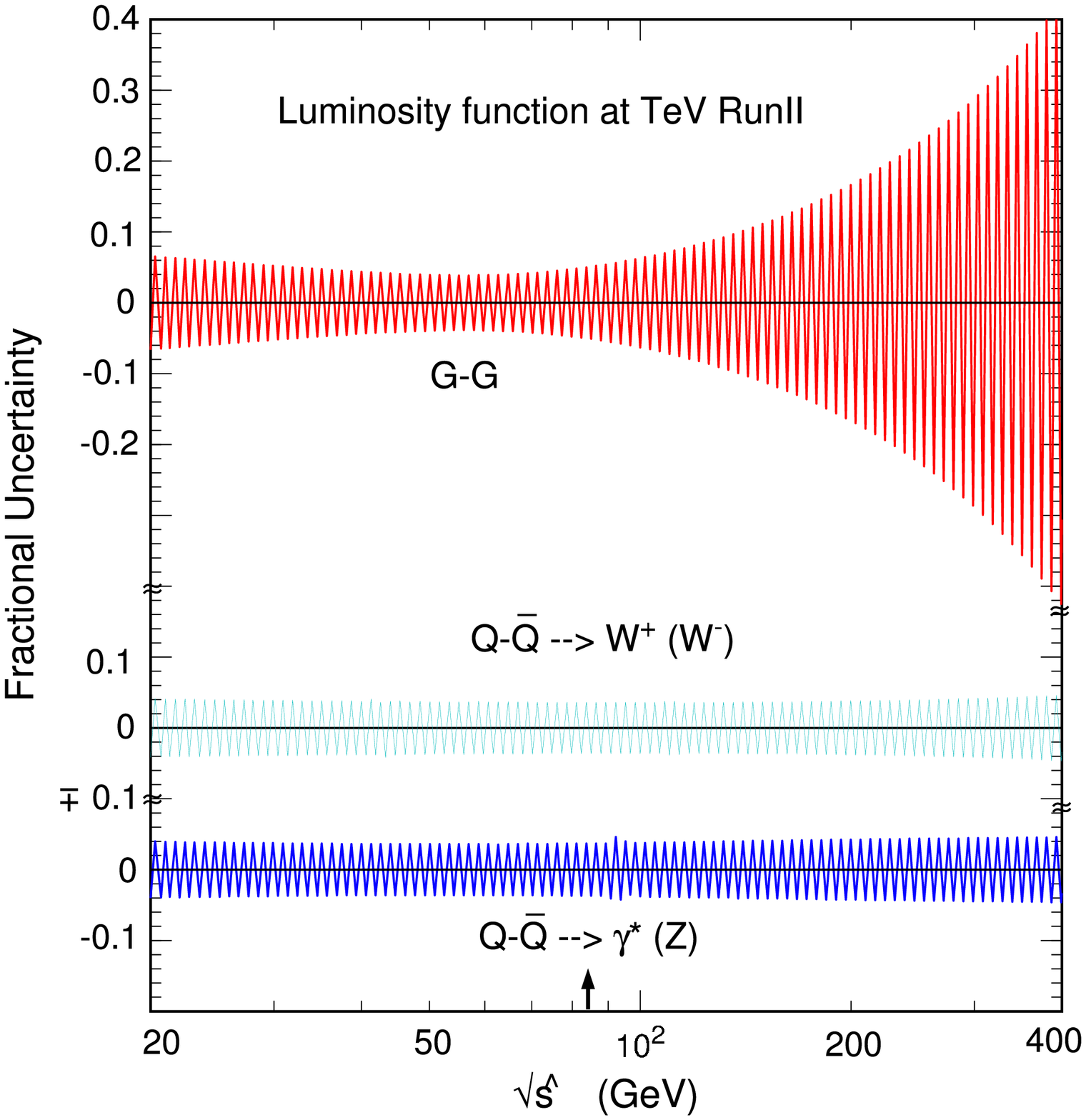}} \hfill
\raisebox{2ex}{
\resizebox{0.47\textwidth}{!}{\includegraphics{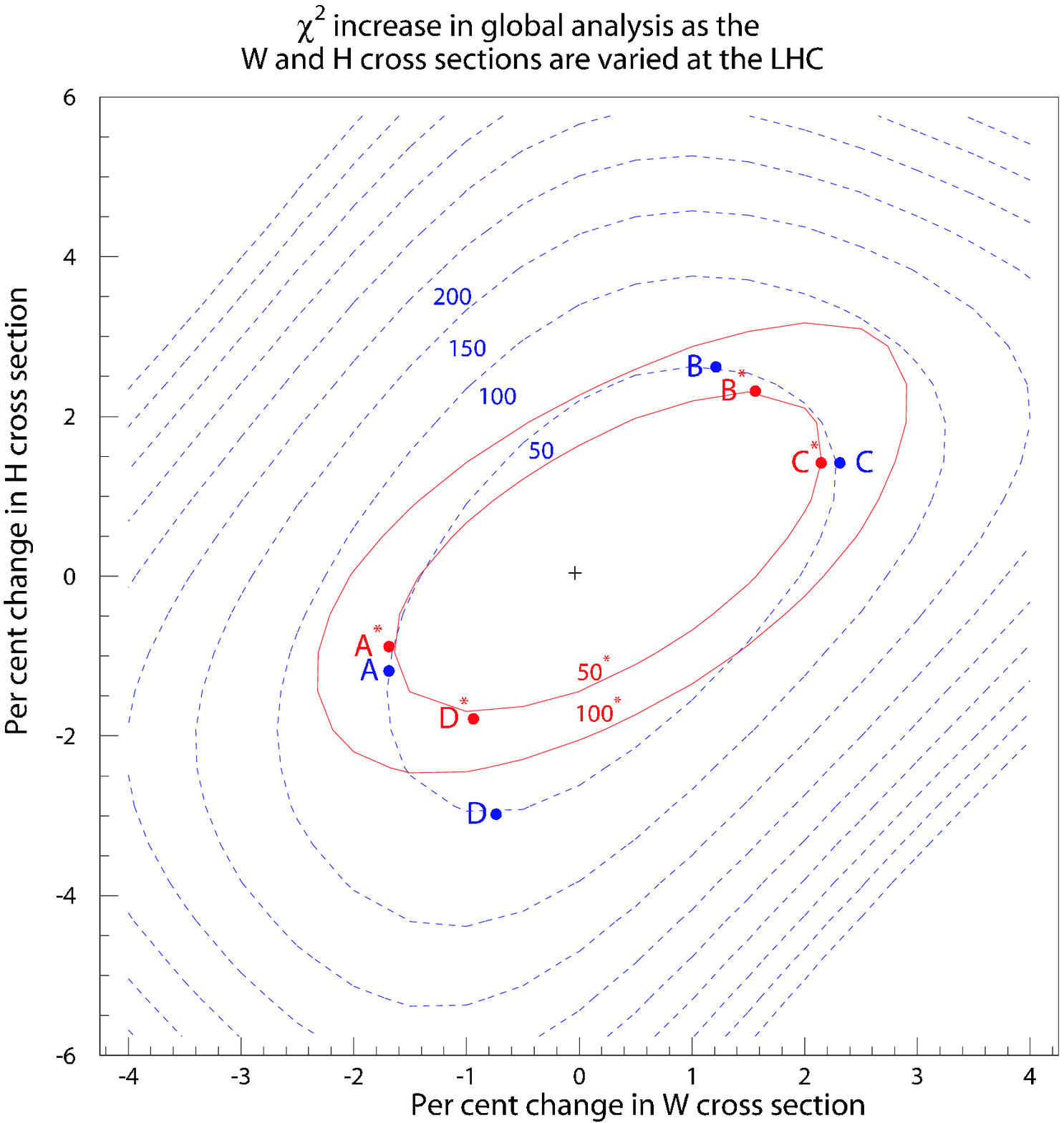}}
}
\caption{Examples of ranges of predictions at Tevatron Run II and LHC associated
with uncertainties of PDFs due to experimental input. See text.}
\label{fig:Unc}
\vspace{-3ex}
\end{figure}
}
\begin{document}

\title{Status of Global QCD Analysis and \\
the Parton Structure of the Nucleon%
 }
\author{Wu-Ki Tung}

\address{Michigan State University, E. Lansing, MI, USA\\
E-mail: Tung@pa.msu.edu }

\maketitle

\abstracts{ The current status of global QCD analysis of parton distribution
functions of the nucleon is reviewed. Recent progress made in determining
various features of the parton structure of the nucleon, as well as
outstanding open questions are discussed. These include: the small-$x$ and
large-$x$ behavior of the partons, particularly the gluon; the
differentiation of $u$ and $d$ quarks; the strangeness sea ($s+\bar{s}$),
the strangeness asymmetry ($s-\bar{s}$); and the heavy quark distributions
$c$ and $b$. Important issues about assessing the uncertainties of
parton distributions and their physical predictions are considered. These
developments are all critical for the physics programs of HERA II, Tevatron Run
II, RHIC, and LHC.}


This talk is devoted to a review of the present status of global QCD analysis
of parton distribution functions (PDFs) of the nucleon. Progress on global
QCD analysis of PDFs depends on continued advancement in experimental inputs
from a variety of hard processes, in theoretical development (such as
higher-order calculations and resummation techniques), and in analysis
methodology (such as statistical techniques and uncertainty assessment).
Significant developments on all these fronts have been taking place. Since
both precision standard model (SM) phenomenology and beyond SM ``new
physics'' searches depend on reliable knowledge of PDFs, continued progress
on global QCD analysis of PDFs is vital for the physics programs of HERA II,
Tevatron Run II, RHIC, and LHC.

2004 marks the 20th anniversary of two landmark works in the field of PDF
analysis---that of Duke-Owens \cite{DuOw} and Eichten-Hinchliffe-Lane-Quigg %
\cite{ehlq}. It seems appropriate to present this review from the historical
perspective, assessing the impressive progress that has been made during the
intervening years, as well as emphasizing the many open problems that remain
to be addressed and resolved. We shall see that in order to meet the challenges
of collider physics in the coming decade, the unfinished tasks in global QCD
analysis continue to be extremely demanding on all fronts.

Due to space limitations, only a few key references are given. Detailed
references to the original literature can be found in various review articles
and on the web pages of the experimental and PDF analysis groups to be
mentioned in the ensuing sections.

\section{Progress in Global QCD Analysis}



Historically, structure functions in deep inelastic scattering (DIS) of
leptons on hadrons, measured in fixed target experiments, provided the first
significant experimental inputs to PDF analysis. They allowed the first
determination of the $u$, $d$ and singlet quark distributions, as well as the
general behavior of the gluon distributions, indirectly through QCD evolution
and sum rules. Lepton-pair production in hadron collisions (Drell-Yan
process) provided dramatic confirmation of the universality of these PDFs,
and furnished critical constraints on the sea quark distributions $\bar{u}$
and $\bar{d}$, thus allowing a meaningful separation of the valance
distributions $u_{\mathrm{val}}$, $d_{\mathrm{val}}$ from the (total) $u$,
$d$ distributions.

Successive generations of DIS experiments, with a variety of lepton beams ($%
e^{\pm },u^{\pm },\nu ,\bar{\nu}$) and hadron targets ($p,D,A$), allowed
better separation of $d$ from $u$ and the valence from the sea distributions.
The ($\bar{d}-\bar{u}$) difference was found to be non-zero, i.e.\ the sea
partons are not flavor-blind.  (This vitiated the Gottfried sum rule.)  The
measurement of the $pD$ to $pp$ Drell-Yan cross section ratio provided
critical information that dramatically altered the behavior of the extracted
$\bar{d}/\bar{u}$ ratio of parton distributions. Complementary to this, the
measurement of the forward/backward asymmetry of the lepton rapidity
distribution in $W$-production (or, equivalently, the $W^{+}/W^{-}$ratio) at
the Tevatron provided valuable constraints on the $d/u$ ratio.

The HERA experiments, H1 and ZEUS, greatly expanded the kinematic range over
which the DIS structure functions are measured, especially for small $x$, and
elevated the precision of these measurements to unprecedented levels. These
advances not only pinned down certain combinations of the quark
distributions to a high degree of accuracy, but also allowed better
determination of the gluon distribution $g(x,Q)$ through the much improved
measurement of the $Q$-dependence of the structure functions. Combined with
direct constraints provided by inclusive jet production measurements in
hadron-hadron collisions (by the CDF and D0 experiments) at medium to large
$x$, we now have a far better picture of the gluon distribution than a few
years ago. Nonetheless, the relative uncertainty on $g(x,Q)$ compared to the
quark distributions is still large.

At the highest $Q$ values reached at HERA, structure functions due to $W^{\pm
}$ (charged current) exchange and $Z$-$\gamma $ interference (neutral
current) have also been measured. They provide dramatic confirmation of the
SM formalism, as well as new opportunities to cleanly differentiate the $u$
and $d$ quarks. (Current analyses, based on DIS on deuteron or nuclear
targets, are sensitive to model-dependent nuclear corrections). However, due
to the small cross section at large $Q$, the accuracy of current measurements is
limited. So far, one can only verify that existing PDFs agree with available
high $Q$ CC and NC HERA data.

The differentiation of strange quark distributions from non-strange ones can,
in principle, be obtained as differences in totally inclusive structure
functions from neutral-current (NC) and charged-current (CC) DIS experiments.
But the systematic uncertainties for such analysis is too large to make this
method reliable. Instead, one has to rely on the semi-inclusive measurement
of charm production in neutrino (and anti-neutrino) scattering. All evidence
from these measurements points to suppressed $s$ and $\bar{s}$ distributions
inside the nucleon compared to the non-strangeness sea quarks [say, $\frac{1%
}{2}(\bar{u}+\bar{d})$] by a factor of roughly $\frac{1}{2}$ at the
confinement scale of order $Q\sim 1$ GeV. This result has been incorporated
in existing global analyses as an overall constant factor. More
detailed analysis of the strangeness sector, including the shapes of $s$ and
$\bar{s}$ distributions and possible differences between them has just begun
recently, as will be described in Sec.\ref{sec:histsur}.

In addition to the ``light partons,'' we also need to understand the partonic
structure of the nucleon in the heavy flavor sectors, particularly $c$ and $b
$---for their intrinsic importance, as well as their influence on processes
such as top and Higgs production. All phenomenological work, so far, assumes
``radiatively generated'' heavy flavors---the ($c,b,t$) partons are generated
by QCD evolution without a non-perturbative component at the respective heavy
flavor threshold. This is not a satisfactory situation because the ``heavy
flavor threshold'' is not a well-defined PQCD concept (any value of the same
order of magnitude as the heavy quark mass is acceptable). A zero value for
the heavy quark distribution at one threshold value would imply non-zero
values at other possible choices for the threshold.  Nonetheless, the PQCD
formalism for treating heavy flavor partons has developed steadily in
the last ten years. Quantitative, systematic phenomenological work is only at
an early development stage. To make substantial progress, more data on heavy
flavor production from a variety of processes will be needed.

\subsection*{Historical Development}

\label{sec:histsur}

To gain a historical perspective on the progress of our knowledge of the
parton structure of the nucleon, we start with Table I, which shows how a
sequence of representative historical PDFs fare when they are compared with
the currently available high precision data.  This comparison is meant to
show the broad historical trend, thereby to give a semi-quantitative measure
of the progress that has been made through these years.  To this end, we
convolute these PDF sets, in turn, with the \emph{same} set of hard cross
sections (that used in the CTEQ global analysis); compute the $\chi ^{2}$
values for the various processes; then examine the results. This procedure
inevitably involves some mismatch between the PDFs and the hard cross sections
for PDF sets obtained by other groups (such as MRST and GRV) because of
differences in the choice of schemes and prescriptions for calculating hard
cross sections. To emphasize the historical trend only, we group the PDF sets
by periods and average over the sets from various groups in the last column,
which carries the main message. (The numbers in adjacent rows within each
group should not be taken literally as measures of relative merit.) \Table By
this account, the accuracy of PDFs has improved steadily, by a factor of ten
over these years. The impact of second generation fixed-target DIS
experiments (BCDMS, NMC, CCFR), $e$-$p$ collider experiments (H1, ZEUS), and
the addition of DY-W production experiments, and jet inclusive production
experiments (CDF, D0), recounted in the first part of this section, all contribute
noticeably in the progression of improvements. \ We next survey the influence
of these developments on the behavior of the various PDFs themselves.

\vspace{2ex}

\noindent\textbf{The $u$ and $d$ distributions:}

The two graphs in Fig.(\ref{fig:udquarks}) show the historical evolution of
the $u$ and $d$ distributions at $Q^{2}=10$ GeV$^{2}$.
\udquarks
We see that
improved data have caused considerable changes in the functions $u(x,Q)$ and
$d(x,Q)$ over the years, particularly at small-$x$ (as a direct consequence
of the advent of the HERA data since the mid-1990's). Combined high
precision fixed-target and HERA collider data now cover roughly the $x$ range of ($%
10^{-5},0.75$). These result in rather well-determined and stable $u$ and $d$
distributions in most recent global analyses. The remaining uncertainties
concern mainly the large $x$ behavior, beyond the measured range,
particularly the $d/u$ ratio. (Cf.\ Sec.\ref{sec:uncert}.) The discrimination
between these two flavors is currently hampered by the dependence on unknown
nuclear effects associated with the necessity to use DIS data on deuteron
targets. (See below.)

\vspace{2ex}

\noindent \textbf{The gluon distribution:}

The historical evolution of the gluon distribution, shown in Fig.(\ref%
{fig:GluAB}a), is more interesting. For reasons mentioned earlier in this section,
the constraints on $g(x,Q)$ are much looser than for $u$ and
$d$. Hence, the first determinations of the gluon distribution varied over a
wide range. The initial HERA data forced a much steeper rise of
$g(x,Q)$ toward small-$x$, similar to the quark distributions. Notice, however, the more
recent global analyses all have a much more moderate rise, or, in some
cases, even a fall, in the small-$x$ behavior of the gluon. An important
contributing factor for this turn-around is the indirect effect of the
inclusion of single-jet inclusive production data from the Tevatron. These
data favor a larger $g(x,Q)$ at high-$x$, which takes away gluons at
small-$x$ because of the overall momentum sum rule! The range over which the
gluon distribution has developed over these years shows vividly both how
global QCD analysis has been evolving, and how much further we need to go to
determine the parton structure of the nucleon with confidence. \GluAB

The last point is reinforced by Fig.(\ref{fig:GluAB}b), which shows the
current uncertainty of the gluon distribution (due to experimental input only)
estimated by the CTEQ analysis \cite{cteq6}. (Cf.\ Sec.\ref{sec:uncert}.) The
fractional uncertainty is largest at high $x$, where experimental constraints
are scarce. At small $x$, the theoretical uncertainty, not included in this
plot, should be much larger than the band shown here. In fact, for certain
choices of scheme and parametrization, the possibility of having negative
gluons has even been raised \cite{mrst03}. This will be discussed in
Sec.\ref{sec:uncert}.

\vspace{2ex}

\noindent \textbf{The} $\mathbf{d}$ \textbf{to} $\mathbf{u}$ \textbf{ratio:}
The differentiation between $d$ and $u$ quarks relies on the difference
between their couplings in neutral current and charged current
interations, and on comparing data on lepton-proton and lepton-deuteron
scattering experiments. The behavior of the ratio $R_{d/u}=d(x,Q)/u(x,Q)$ as
a function of $x$ at $Q^{2}=10$ GeV$^{2}$ is shown in Fig.(\ref{fig:UdAsym}%
a). The small-$x$ region (say, $x<0.1$) is dominated by the sea quarks. This
will be discussed in the next subsection. The larger-$x$ region ($x>0.1$) is
dominated by the valence quarks, the integral of which must be in the ratio $%
1:2$, reflecting the quark number sum rule for the proton. We see that, aside
from the very early PDF sets, which had no experimental constraints, $d/u$ is
relatively well-determined in the region $0.1$ - $0.3$. The possible range
of variation of this ratio as $x\rightarrow 1$ is considerable---it is even
wider than that shown in this plot, because all curves on this plot result
from parametrizations of the PDFs with similar prejudices with respect to
the $x\rightarrow 1$ behavior. A systematic study of the ``theoretical
uncertainties'' has not yet been done. \UdAsym

\vspace{4ex}

\noindent \textbf{Flavor Asymmetry of  non-strange sea quarks:} $\mathbf{%
\bar{d}}$\textbf{-}$\mathbf{\bar{u}}$

From purely perturbative considerations, early work assumed that sea quarks
are flavor-independent, broken perhaps only by quark mass effects. This
suggested the ansatz $\bar{u}(x,Q)=\bar{d}(x,Q)$ as an initial condition for
QCD evolution. Experimental results soon proved that this conjecture is not
correct. Non-perturbative physics at the confinement scale is far more
subtle. The isospin asymmetry of the sea, as measured by $R_{\bar{d}\bar{u}}=%
\frac{\bar{d}(x,Q)-\bar{u}(x,Q)}{\bar{d}(x,Q)+\bar{u}(x,Q)}$ at $Q^{2}=10$
GeV$^{2}$, is shown in Fig.(\ref{fig:UdAsym}b) for the same historical PDF
sets as before. Some very early ones assumed $R=0$, hence lie on the x-axis
of the plot. Subsequent ones, with input from lepton-proton and lepton-deuteron
scattering experiments (mainly in the small-$x$ region that is sensitive to
the sea quarks), correspond to curves that stay on the upper half plane. It
turned out, however, the large-$x$ rise was only a consequence of
extrapolation. After the measurement of the ratio of DY cross sections, $%
\sigma _{pD}/\sigma _{pp},$ became available, particularly from the E866
experiment at Fermilab, the ratio $R$ was found to turn around and decrease
with $x$ beyond $x=0.2$. This plot again underlines the importance of having
the most relevant experimental constraints for each feature of the
non-perturbative PDFs. Without all the necessary experimental input, even the
most sophisticated analysis can yield quite unphysical results, as is
illustrated by the top curve (black) obtained by Alekhin in a NNLO analysis
\cite{alekhin} that does not include DY data.

\vspace{2ex}

\noindent \textbf{Flavor SU(3) Asymmetry of sea quarks: strange vs.
non-strange: } The ``naive'' expectation of flavor SU(3) symmetry, $s=\bar{s}%
=\bar{d}=\bar{u}$, assumed in some early PDF studies, is clearly
unrealistic: the strange quark mass alone would induce a difference between (%
$s,\bar{s}$) and, say, $(\bar{u}+\bar{d})/2$. As mentioned earlier,
experimental evidence suggests that the ratio
$R_{s+}\equiv \frac{s+s}{\bar{u}+\bar{d}}$, is of the order 0.5 at a scale of
1-2 GeV. Up to now, this ratio is mainly enforced as a constraint in global
analyses, rather than as the result of actual fitting to data, because the
relevant data have not been presented in a form suitable for global analysis.
We show the actual values of $R_{s+}$ for the various PDF sets used as
examples in Fig.(\ref{fig:StrAsym}a).  The somewhat different values of
$R_{s+}$ for various sets seen on this plot is more due to the minor
variation on how and where the condition $R_{s+}\sim 0.5$ is implemented,
rather than on hard experimental input. Thus, there is no regular pattern in
terms of the historical development (other than the departure from the naive
perturbative $R_{s+}=1$ assumption).

\StrAsym

\noindent \textbf{Strangeness Charge Asymmetry:} In the absence of any strong
theoretical argument or experimental evidence to the contrary, the strange
and anti-strange quark distributions, $s(x,Q)$ and $\bar{s}(x,Q)$, have
traditionally been assumed to be equal to each other in all widely available
PDFs. Recent developments, particularly in relation to the NuTeV anomaly,
have motivated a re-examination of this assumption. In a phenomenological
study performed in the global analysis context \cite{StrAsym}, analysis of
the CCFR-NuTeV data on dimuon (charm) production in neutrino and
anti-neutrino scattering gave preliminary evidence of a positive
non-perturbative strangeness asymmetry in terms of the first moment integral
$[S^{-}]\equiv \int_{0}^{1}x[s(x)-\bar{s}(x)]dx$. In
Fig.(\ref{fig:StrAsym}b), we show the result of this analysis, along with
earlier results that have known problems. (See \cite{StrAsym}.) From
perturbation theory, it has been pointed out that the 3-loop QCD evolution
kernel generates a non-zero (negative) strangeness asymmetry, even if one
starts with a symmetric non-perturbative input to the analysis. The situation
is clearly unsettled at the moment.

\vspace{2ex}

\noindent \textbf{Possible Isospin Violation in the Parton Structure of the
Nucleon:} Interest in possible explanations of the NuTeV anomaly has also
motivated the study of the possibility that $u_{\mathrm{proton}}(x,Q)\neq d_{%
\mathrm{neutron}}(x,Q)$. Experimental constraints on this effect are very
weak, even without taking into account the large uncertainties about nuclear
corrections that are needed to measure neutron structure functions
\cite{mrst03}. Theoretically, it has been pointed out that isospin violation
in PDFs arises naturally when one tries to include electroweak corrections in
the global QCD analysis: the evolution equations of PDFs will then include
photon distributions of the nucleon; $u_{\mathrm{proton}}(x,Q)$ and
$d_{\mathrm{neutron}}(x,Q)$ distributions will evolve differently due to their
different electric charge. This effect has been studied; it is small, as
expected.

\vspace{2ex}

\noindent \textbf{Heavy Quark distributions--Charm and Bottom:} Although
there has been much discussion about the physical processes involving heavy
quark production in the literature, there is as yet not very much reliable
information on the heavy flavor parton distributions. Of the heavy quarks $%
c,b,$ and $t,$ only the $c$ and $b$ quark-partons participate actively in
PQCD calculations of high energy processes for physical processes at energy
scales even up to LHC. The definition of heavy quark partons is even more
scheme-dependent than that of light quark flavors. In the so-called (fixed)
3-flavor scheme, there are, by definition, no heavy quark partons at all;
whereas in the (fixed) 4-flavor scheme, there is a charm distribution, but no
bottom distribution. These schemes are of use only for limited energy ranges.
In recent years, a consensus has emerged that the variable-flavor number
scheme, which is a generalization of the conventional msbar zero-mass parton
scheme, is the appropriate one to use for calculations that cover a wide
range of $Q$. If one assumes that there is no non-perturbative heavy flavor
content in the nucleon, then heavy flavor parton distribution functions $c$
and $b$ are ``radiatively generated'' by QCD evolution from their respective
thresholds. This is the assumption used in practically all existing global
analyses of PDFs. Unfortunately, the concept of radiatively generated heavy
quark partons is not well-defined, since the location of ``heavy quark
threshold'' for a given flavor is itself ambiguous: it can be any value of
the same order of magnitude as the heavy quark mass or the physical heavy
flavor particles.

\charm Fig.(\ref{fig:charm}) shows the charm distribution at $Q^{2}=10$ GeV$%
^{2}$ for the various PDF sets. The small-$x$ behavior of the radiatively
generated distribution has changed rather dramatically over time, mainly as
the indirect consequence of the change of small-$x$ behavior of the light
partons, driven by the HERA data. For reasons mentioned above, the actual
behavior of $c(x,Q)$ remains largely unexplored. Both new data and new
theoretical assumptions could change these results rather drastically.

\section{Open Issues in Global QCD Analysis}

\label{sec:open}

The above survey of various aspects of the parton structure of the nucleon
amply illustrates that, although a lot of progress has been made over the
last 20 years, many features of the PDFs still are uncertain, or unknown.
These open issues are summarized here.

\mylis[$\bullet$]{ \item The gluon distribution: both the small $x$ and
large-$x$ behavior of the gluon distribution $g(x,Q)$ are still subject to
large uncertainties. Two current results are worth further investigation.

\mylis[$\diamond$]{ \item The MRST global analysis favors a negative gluon at
small-$x$ at a momentum scale of $1$ GeV $<Q<2$ GeV. Is this ``required'' by
current experimental input?  Is this behavior ``natural'' theoretically?

\item The CTEQ6 gluon distributions at large $x$ are higher than in most
other PDF sets; $g(x,Q_{0})$ falls off as $(1-x)^{\alpha }$ with a power
$\alpha $ that is slightly smaller than the power for the valence quark. This
is driven by the inclusive jet data used in the analysis. But is it
theoretically ``natural''? }

\item The $R_{d/u}$ ratio: should this ratio approach $0$, $\frac{1}{4}$,
$\frac{3}{7}$, or some other number, as $x\rightarrow 1$? How do we determine
the large-$x$ behavior of this ratio phenomenologically? Is there a reliable
way to apply deuteron corrections to data obtained with deuteron target,
which can significantly affect the analysis?

\item The strange quark distribution: the strangeness sector of PDFs is most
conveniently studied in terms of $s_{\pm }=s\pm s$, which correspond to
different quantum numbers, and embody different underlying physics.

\mylis[$\diamond$]{ \item The symmetric combination $s_{+}(x,Q)$ is easier to
study since it can be probed by charm production in DIS experiments using
combined neutrino and anti-neutrino beams. After many years of effort,
however, the uncertainty on $s_{+}$ is still quite large. For instance, aside
from the relative suppression of strange compared to non-strange sea, we
cannot yet be certain whether $s_{+}(x,Q_{0})$ can be sufficiently
represented by the same functional form as, say
$\bar{u}(x,Q_{0})+\bar{d}(x,Q_{0}),$ with only a normalization factor $\kappa
$. A detailed NLO global analysis, including the recent CCFR-NuTeV dimuon
production data, can possibly yield an improved knowledge on
$R_{s+}(x,Q)\equiv \frac{s+s}{\bar{u}+\bar{d}},$ including a quantitative
estimate of the uncertainties. This remains to be carried out.

\item The antisymmetric combination $s_{-}(x,Q)=s(x,Q)-\bar{s}(x,Q)$ is
harder to determine, because it requires the measurement of the difference
between neutrino and anti-neutrino cross sections with charm final states.
The status of this determination was briefly discussed in the previous
section. (Cf. Ref.\cite{StrAsym}) In practice, it is convenient to
parametrize the ratio $R_\mathrm{StrAsy}(x,Q_{0})\equiv
\frac{s(x)-\bar{s}(x)}{s(x)+\bar{s}(x)}\displaystyle|_{Q_{0}}$which must
satisfy the constraint $\left\vert R_\mathrm{StrAsy}\right\vert \leq 1$ in
order to ensure positivity of the distributions $s$ and $\bar{s}$. A full NLO
global analysis of this quantity is also currently underway, in conjunction
with the $s_{+}$ study described above. }

\item Isospin Violation in the Parton Structure of the Nucleon: \\
(See the previous section.)

\item Heavy Quark Distributions: (See the previous section.)
} 

\section{Uncertainties of Parton Distributions}

\label{sec:uncert}

In parallel with the determination of ever improving ``best-fit" PDFs, an
equally important front in global analysis has been opened in recent
years---the development of quantifiable uncertainties on the PDFs and their
physical predictions.  Several groups have carried out extensive studies with
different techniques and emphases. Much progress has been made; many useful
results have been obtained; but there are, as yet, no unambiguous
conclusions. The basic problem lies with the complexity of the global
analysis that (i) utilizes results from many experiments on a variety of
physical processes, with diverse characteristics and errors, and often not
mutually compatible according to textbook statistics; (ii) may be sensitive
to many theoretical uncertainties that cannot yet be quantified; and (iii)
can depend on the choice of parametrization of the non-perturbative functions
used in the analysis. Individually and collectively, these factors render a
\emph{rigorous} approach to error analysis untenable.

As an illustration of point (i), we briefly describe results on a study of
the uncertainty of the W production cross section at the Tevatron due to
known experimental errors on the input data sets, conducted by the CTEQ group
\cite{cteq6} using the Lagrange Multiplier method they proposed. First, we
obtain a series of PDFs that provide best fits to the global data, but
constrained to yield a range of possible values of $\sigma _{W}$ at the
Tevatron around the CTEQ6M value (which corresponds to the least overall
$\chi ^{2}$ by definition). Then, we evaluate the individual $\chi ^{2}$ of
each experimental data set to gauge the consistency between the data sets, as
well as to assess sensible ways to quantify the overall uncertainty of the
prediction on $\sigma _{W}$ due to the input experimental uncertainties. The
results are shown in the two plots of Fig.(\ref{fig:wXsecErr}). \wXsecErr The
horizontal axes correspond to the values of $\sigma _{W}.$ For each of the 15
input experimental data sets, a best-fit value and a range is shown. These
are arranged vertically, in no particular order. On the left plot, each range
corresponds to a $\Delta\chi^2=1$ error due to that experiment; while on the
right plot, it corresponds to a ``90\% confidence level'' (cumulative
distribution function of the $\chi ^{2}$ normalized to the best fit). We see
clearly: (i) if  a $\Delta\chi^2=1$ error criterion is strictly enforced,
then there is no common value for the predicted $\sigma _{W}$ (or,
equivalently, some of the data sets must be deemed mutually incompatible);
but (ii) within the 90\% confidence level range, there is a common range for
$\sigma _{W}$ that spans the values indicated by the dashed vertical lines.

Faced with the problem of nominally incompatible data sets (which is common in
combined analysis of data from diverse experiments, e.g.\ PDG work),
subjective assumptions and compromise measures are necessary to obtain
sensible results. Several approaches have been followed by the different
global analysis groups. CTEQ uses the ansatz that the range of uncertainty
indicated in Fig.(\ref{fig:wXsecErr}b) represents a 90\% C.L.~uncertainty on
$\sigma _{W}$; and, in general, characterizes the PDF uncertainties by using
similar criteria along 20 orthonormal eigenvector directions in the PDF
parameter space, using an improved Hessian method.\footnote{%
In terms of the total $\chi ^{2}$ of the global data sets, consisting of $%
\sim 2000$ data points in current analysis, this range corresponds to $%
\Delta \chi ^{2}\sim 100.$ There is no a priori significance to this number,
since the global $\chi ^{2}$ used in this context only represents a broad
measure of goodness-of-fit; it does not have rigorous statistical
significance. As data increase in quantity and quality, the equivalent $%
\Delta \chi ^{2}$ value will change. Similarly, when applied to a different
observable or set of input data, the number will vary.} MRST has adopted the
same approach \cite{mrst02}, albeit choosing a slightly narrower range of the
uncertainty. The H1 and ZEUS PDF analysis groups also adopt similar methods,
but, by restricting the input data sets to DIS experiments only, apply their
own definition of the range.~\cite{Mandy} The important fact is that these
different groups (all using the leading twist PQCD formalism) arrive at quite
comparable results, both for the PDFs and for the magnitude of the error
bands, even if some details are different because of the variations in
experimental and theoretical inputs.

With this approach, both CTEQ and MRST have been able to make estimates on
future measurements. Two examples are given in Fig.(\ref{fig:Unc}). \Unc
Fig.(\ref{fig:Unc}a) shows fractional uncertainties in the predicted
$q\bar{q}$ and $GG$ parton luminosity functions as a function of
$\sqrt{\hat{s}}$ at the Tevatron energy obtained by CTEQ, from which the
values and the uncertainty ranges of a variety of physical processes, both in
the SM and beyond, can be obtained. We see the considerable uncertainties
associated with the gluon-gluon luminosity at large $x$. Fig.(\ref{fig:Unc}b)
shows contours of increasing $\chi ^{2}$ in the $\sigma _{W}$-$\sigma _{H}$
plane due to PDF uncertainties at LHC obtained by MRST. Theoretical
uncertainties are not included in either plot.

A different approach is followed by Alekhin \cite{alekhin}. The experimental
input is restricted to DIS experiments only, and the theoretical framework is
broadened to include higher-twist effects, among others, in order to better
accommodate the different data sets. A consistent fit is then achieved in the
strict statistical sense; and the uncertainty range is defined according to
the classic $\Delta \chi ^{2}=1$ criterion. However, by forgoing the critical
experimental constraints provided by Drell-Yan and inclusive jet production
data, the determination of the PDFs can not be complete. Applying the Alekhin
PDFs to the available DY data sets (E605, CDF W-asymmetry, E866), one obtains
a $\chi ^{2}$ of 892 for 145 data points---a clear indication that vital
information is missing on certain aspects of PDFs. This can be seen in the
plot of $\frac{\bar{d}-\bar{u}}{\bar{d}+\bar{u}}$ shown in
Fig.(\ref{fig:UdAsym}b), where the Alekhin curve is similar to earlier PDFs
that did not include DY asymmetry data sets, but is completely different from
other modern PDFs that do fit the DY data. Under these circumstances, one
might
ask, what is the use of these PDF uncertainties defined by the textbook $%
\Delta \chi ^{2}=1$ rule? Giele \textit{et al} \cite{giele} also emphasize a
rigorous statistical approach, using the more general likelihood method.
Within the leading twist PQCD approach, this leads to acceptable results only
if one restricts the input experimental data sets to one or few DIS sets.
Thus, depending on which subsets of data are used, one gets many predictions
on physical
quantities (such as $\sigma _{W}$) with ``1$\sigma $%
-error'' ranges, which do not overlap with each other. This leaves unanswered
the important question: ``What is the best estimate of current uncertainty,
given all available experiments?''.

Thus, the underlying facts seen by all groups are consistent with each other;
the differences lie in the emphases placed to cope with these facts. In
principle, all methods are equivalent: in an ideal world where all
experiments came up with textbook-like errors, they would yield the same
results. In reality, in the complex world of global analysis, the results
appear different or non-existent (if strict criteria are applied), depending
on subjective judgements made in placing the emphases. This state of affairs
requires the users of PDFs to be well-informed about the nature of the
``uncertainties'' provided by the various global analysis groups; and then to
use these judiciously according to their own (subjective) judgement.
Unfortunate it may be, but there is no ``1-$\sigma $ PDF error'' that can be
defended scientifically on all accounts. This points to the need for
continued hard work, both on the experimental and theoretical fronts, in
order to improve the situation, and to reduce the ambiguities described
above. To this end, the physics programs of HERA II, Tevatron Run II, as well
as several fixed-target experiments, can make important contributions in the
immediate future. Through these efforts, the PDFs and their uncertainties
will certainly be better known when the LHC comes on line. This will lead to
better predictions on both SM and new physics processes, hence improve the
potential for all discoveries. In addition, the high reaches of LHC, both in
energy range and in statistics, will provide additional constraints on PDFs,
hence allow even better determination of the parton structure of the nucleon.

\section{Other important topics relating to global analysis}

Due to space limitation, many areas of active work relevant for global QCD
analysis cannot be included in this survey. These include: (i) the
completion of the NNLO (3-loop) calculation of the QCD evolution kernel;
(ii) important advances in the calculation of NLO and NNLO hard cross
sections for a variety of physical processes; (iii) inclusion of electroweak
effects in the QCD global analysis; (iv) estimates of theoretical
uncertainties, with implications on the stability of the NLO global
analysis; and many more. They are covered elsewhere in the proceedings,
either in the parallel sessions or in the concluding plenary sessions.

\section*{Acknowledgements}

Much of the material reported in this review is based on CTEQ work,
particularly in collaboration with my MSU colleagues Joey Huston, Jon Pumplin
and Dan Stump, and on published results of other QCD analysis groups. The
perspectives expressed here are borne out of fruitful discussions with
members of CTEQ, MRST, the PDF analysis groups of H1 and ZEUS, S.~Alekhin,
W.~Giele, and many others in the global QCD analysis community.


\begin{thebibliography}{9}
\bibitem{DuOw}
D.~W.~Duke and J.~F.~Owens,
Phys.\ Rev.\ D {\bf 30}, 49 (1984).
%
\bibitem{ehlq}
E.~Eichten, I.~Hinchliffe, K.~D.~Lane and C.~Quigg,
Rev.\ Mod.\ Phys.\  {\bf 56}, 579 (1984) [Addendum-ibid.\  {\bf 58}, 1065
(1986)].
%
\bibitem{cteq6}
J.~Pumplin, D.~R.~Stump, J.~Huston, H.~L.~Lai, P.~Nadolsky and W.~K.~Tung,
JHEP {\bf 0207}, 012 (2002) [arXiv:hep-ph/0201195], and references therein.
%
\bibitem{mrst03}
A.~D.~Martin, R.~G.~Roberts, W.~J.~Stirling and R.~S.~Thorne,
Eur.\ Phys.\ J.\ C {\bf 35}, 325 (2004) [arXiv:hep-ph/0308087].
%
\bibitem{alekhin}
S.~Alekhin,
Phys.\ Rev.\ D {\bf 68}, 014002 (2003) [arXiv:hep-ph/0211096], and references
therein.
%
\bibitem{StrAsym}
S.~Kretzer, F.~Olness, J.~Pumplin, D.~Stump, W.~K.~Tung and M.~H.~Reno,
Phys.\ Rev.\ Lett.\  {\bf 93}, 041802 (2004) [arXiv:hep-ph/0312322];
%
F.~Olness {\it et al.},
arXiv:hep-ph/0312323.
\bibitem{mrst02}
A.~D.~Martin, R.~G.~Roberts, W.~J.~Stirling and R.~S.~Thorne,
Eur.\ Phys.\ J.\ C {\bf 28}, 455 (2003) [arXiv:hep-ph/0211080], and
references therein.
%
\bibitem{Mandy}
A.~M.~Cooper-Sarkar,
J.\ Phys.\ G {\bf 28}, 2669 (2002) [arXiv:hep-ph/0205153];
\\
also http://www-zeuthen.desy.de/~moch/heralhc/gwenlan-coopersarkar.pdf

\bibitem{giele}
W.~T.~Giele and S.~Keller,
Phys.\ Rev.\ D {\bf 58}, 094023 (1998) [arXiv:hep-ph/9803393];
%
W.~T.~Giele, S.~A.~Keller and D.~A.~Kosower,
arXiv:hep-ph/0104052 (unpublished).
%
\end{thebibliography}
\end{document}